\documentstyle[12pt,lathuile]{article}
\def\Xn{\tilde{\chi}_1^0}
\def\Xpm{\tilde{\chi}_1^{\pm}}
\def\Ohsq{\Omega_{\tilde{\chi}_1^0}h^2}
\def\SXp{\sigma_{\tilde{\chi}_1^0-p}}
\def\TB{\tan \beta}
\def\BSG{b \rightarrow s \gamma}
\input epsf.tex
\def\DESepsf(#1 width #2){\epsfxsize=#2 \epsfbox{#1}}
\def\Journal#1#2#3#4{{#1} {\bf #2}, #3 (#4)}

\def\NPB{{Nucl. Phys.} B}
\def\PLB{{Phys. Lett.} B}
\def\PRL{Phys. Rev. Lett.}
\def\PRD{{Phys. Rev.} D}
\def\ZPC{{Z. Phys.} C}
\def\AP{Astropart. Phys.}

\begin{document}
\title{ 
EFFECTS OF THE MUON $g-2$ ANOMALY ON \\
DARK MATTER AND ACCELERATOR PHYSICS
}
\author{R. Arnowitt, B. Dutta , B. Hu, and Y. Santoso       \\
{\em Center For Theoretical Physics, Department of Physics, Texas A\&M 
University,} \\ {\em College Station, TX 77843-4242, USA} }
\maketitle
\baselineskip=14.5pt
\begin{abstract}
The effect of the recently observed $2.6\,\sigma$ deviation of the muon 
anomalous magnetic moment ($a_{\mu} = (g_{\mu} -2)/2$) from its Standard Model 
prediction is examined  within the framework of supergravity models with 
grand unification and R parity invariance. The constraints of the Higgs 
mass bounds, the $\BSG$ bounds (including the large $\tan\beta$ NLO 
corrections) and the cosmological relic density of light neutralinos 
(including all slepton neutralino coannihilation effects) are included in 
the analysis. For universal soft breaking, the Higgs and $\BSG$ 
bounds puts a lower bound $m_{1/2} \stackrel{>}{\sim} 300$ GeV, most of the
parameter space now 
falling in the co-annihilation region. The $2\,\sigma$ lower bound on the 
magnetic moment anomaly places an upper bound of $m_{1/2} \stackrel{<}{\sim}
800$ GeV. It is seen 
that mSUGRA requires that $a_{\mu} \stackrel{<}{\sim} 50 \times 10^{-10}$. One
finds for $m_h > 114$ GeV, 
that $\tan\beta > 5(7)$ for $A_0 = 0(-4 m_{1/2})$ and for $m_h > 120$ GeV, one
has 
$\tan\beta > 15 (10)$ for $A_0 = 0(-4m_{1/2})$. The sparticle spectrum is now
much 
constrained, and the reaches of the Tevatron RUN II, NLC, and LHC for new 
physics discovery are discussed. Dark matter detection rates are examined, 
and it is seen that future detectors now would be able to scan most of the 
parameter space. Models with non-universal soft breaking in the Higgs and 
third generation of squarks and sleptons are exmained , and it is seen that 
a new $Z$ s-channel annihilation of neutralinos in the early universe is 
possible with dark matter detection rates accessible to the next round of 
detectors.  
\end{abstract}
\baselineskip=17pt
\newpage
\section{Introduction}
Recently, the Brookhaven E821 experiment has measured the muon anomalous 
magnetic moment, $a_{\mu} = (g_{\mu} -2)/2$ for the $\mu^+$ to remarkable
accuracy~\cite{b1}:
\begin{equation}
   a_{\mu} = 11 659 202 (14)(6) \times 10^{-10}    \label{eq1} 
\end{equation}
This had led to a $2.6 \, \sigma$ deviation from the prediction of the Standard 
Model (SM):
\begin{equation}
   a_{\mu}^{\rm exp} - a_{\mu}^{\rm SM} = 43 (16) \times 10^{-10} \label{eq2} 
\end{equation}
Eq. (1) arises from an analysis of only about one quarter of the existing 
$\mu^+$ data. Further, the 2001 run for the $\mu^-$ should result in about
three 
times as much data as has been analysed for the $\mu^+$~\cite{b2}. Thus the
statistical 
error in the measurement should decrease by about a factor of $2.5$. In 
addition, the theoretical error in the hadronic vaccum polarization 
contribution should also be reduced (perhaps by a factor of 2) by the 
analysis of the Novosibirsk, DAPHNE and Beijing data. (For an analysis of 
the current theoretical accuracy of the hadronic vacuum polarization
contribution see~\cite{b3}, and for an 
alternate view see~\cite{b4}.) Thus by the beginning of 2002 it should be 
possible to know whether the anomaly in the muon magnetic moment of Eq. 
(2)  is real. If it is indeed real it would imply the discovery of new 
physics. Because of the importance of this possibility, we will here assume 
that the anomaly is real, and analyze some of its consequences.

Supersymmetry offers a possible explanation of a deviation of $a_{\mu}$ from
its 
SM value, and initial calculations of $a_{\mu}$ were made in the early 1980's 
using global supersymmetry~\cite{b5}. However, in unbroken global supersymmetry 
there is a theorem that the total value of $a_{\mu}$ should vanish~\cite{b6},
i.e.
\begin{equation}
     a_{\mu}^{\rm SM} + a_{\mu}^{\rm SUSY} = 0   \label{eq3}
\end{equation}
Thus one needs broken supersymmetry to get a non-zero result, and how to do 
this in a phenomenologically acceptable way within the framework of global 
supersymmetry was problematical. The advent of supergravity (SUGRA), where 
acceptable spontaneous breaking of supersymmetry occurs, led to the 
SUGRA grand unified (GUT) models~\cite{b7}. The first (partial) 
calculation of $a_{\mu}^{\rm SUGRA}$ was done in~\cite{b8}, and the first
complete 
calculation in~\cite{b9}. In the SUGRA GUT models, SUSY breaking at the GUT 
scale $M_G$, triggers the breaking of $SU(2)  \times U(1)$ at the electroweak
scale $M_{\rm EW}$, and hence the SUSY mass scale $M_{\rm SUSY}$ obeys
\begin{equation}
      M_{\rm SUSY} \cong M_{\rm EW} \cong \langle H \rangle	\label{eq4} 
\end{equation}
setting the scale of the SUSY masses to be in the 100 GeV - 1 TeV range, a 
range also needed to avoid the hierarchy problem. This mass scale was 
further supported by the fact that the LEP data is consistent with grand 
unification if the SUSY masses lie in the range $\sim 100$ GeV - 1 TeV.
Further, SUGRA models with R parity invariance have a dark matter candidate, 
the lightest neutralino, $\Xn $, with the astronomically observed
amount of 
relic density when the SUSY masses are in this range. Thus with the mass 
range implied by the SUGRA GUT models, it was possible to predict roughly 
the size of $a_{\mu}^{\rm SUGRA}$, and in fact in~\cite{b9} it was argued that 
$a_{\mu}^{\rm SUGRA}$ 
should become observable when the experiments had a sensitivity of about $20
\times 10^{-10}$, as is the case for Eq. (2).

We consider here, then the effect of $a_{\mu}^{\rm SUGRA}$ for SUGRA GUT models 
possessing R parity invariance. In particular, we consider (1)  the mSUGRA 
model~\cite{b7} with universal soft breaking at $M_G$, and (2) models with 
non-universal scalar masses at $M_G$ for the Higgs bosons and the third 
generation of squarks and sleptons.

\section{Calculational Details}
In the following we impose the current experimental bounds on the SUSY 
parameter space and list now the important ones:
\begin{enumerate}
\item Accelerator bounds: (i) We consider two possible values for the light 
Higgs mass $m_h > 114$ GeV and $m_h > 120$ GeV. The first is the current LEP
bound~\cite{b10} while the second represents a bound that the Tevatron Run IIB could 
achieve, if it does not discover the Higgs boson.  (ii). We assume for the 
$b \rightarrow s \gamma$ branching ratio the roughly $2 \sigma$ range:
\begin{equation}
     1.8 \times 10^{-4} < BR(b \rightarrow s \gamma) < 4.5 \times 10^{-4}
   \label{eq5}
\end{equation}
\item Relic density bounds for the lightest neutralino. We chose here the range
\begin{equation}
     0.025 \leq \Ohsq < 0.25 	\label{eq6}
\end{equation}
where $\Omega_{\Xn} = \rho_{\Xn}/\rho_c$, and
$\rho_c$ is the critical density to close the 
universe: $\rho_c = 3 H_0^2/ 8 \pi G_N$ (and $H_0$ is the Hubble constant, $H_0
= h \; 100$ km/sec Mpc, $G_N$ the Newton constant). The lower bound in Eq. (6)
is smaller
than the conventional value of $\sim 0.1$, but can take into account the
possibility 
that there is more than one species of dark matter. Our results here, 
however, are insensitive to this lower bound and would not change 
significantly if it were raised to 0.05 or 0.1.
\item We use a $2\sigma$ bound on the muon magnetic moment anomaly of Eq. (2):
\begin{equation}
      11 \times 10^{-10} < a_{\mu}^{\rm SUGRA} < 75 \times 10^{-10} \label{eq7}
\end{equation}
\end{enumerate}
As we will see these will combine to put strong constraints on the SUSY 
parameter space and allow us to greatly strengthen the predictions  for the 
neutralino-proton dark matter cross sections $\SXp$, 
as well as determine what parts of the SUSY mass spectrum can be seen at the 
Tevatron Run II, LHC and NLC accelerators.

In order to get accurate results, it is necessary to include a number of 
corrections in the calculation, and we mention here the most important 
ones: (1) The allowed SUSY parameter space is now quite sensitive to the 
Higgs mass, and so an accurate calculation of $m_h$ is important. We include 
here the one and two loop corrections~\cite{b11} and the pole mass corrections. 
The theoretical value of $m_h$ is still uncertain by about 3 GeV, and so the 
experimental bound of $m_h > 114$ GeV will (conservatively) be interpreted as 
a theoretical evaluation of $m_h > 111$ GeV (and similarly $m_h > 120$ GeV as 
$m_h > 117$ GeV). (2) Large $\tan\beta$ NLO corrections to the $b \rightarrow s
\gamma$ decay 
rate~\cite{b12} are included. (3) Loop corrections to $m_b$ and $m_{\tau}$ are
included 
(which are important for large $\tan\beta$). (4) QCD renormalization group 
corrections in going below the scale $M_{\rm SUSY}$  are used. (5) In the relic 
density calculation, all slepton neutralino co-annihilation
effects are included with an analysis valid for large $\tan\beta$. There are
now 
several groups that have carried out this calculation~\cite{b13,b14,b15}, and
we  have checked that they are generally in good agreement.

We have not assumed any specific GUT group relations, such as Yukawa 
unification or proton decay constraints (except for grand unification of 
the gauge coupling constants). Such relations depend on physics beyond the 
GUT scale about which little is known, and need not, for example, hold in 
string models, even when coupling constant unification is required.

\section{mSUGRA Model}
The mSUGRA model~\cite{b7} is the simplest possible model with universal soft 
breaking masses at $M_G$. It depends on four parameters  and one sign: $m_0$,
the 
masses of the scalar particles at $M_G$; $m_{1/2}$, the gaugino masses at $M_G$;
$A_0$, 
the cubic soft breaking mass at $M_G$; $\tan\beta$, the ratio of the Higgs VeVs,
$\langle H_2 \rangle / \langle H_1 \rangle$, at the electroweak scale; and the
sign of $\mu$, the Higgs mixing 
parameter in the superpotential ($W_{\mu} = \mu H_1 H_2$). We allow the above 
parameters to vary over the following ranges: $0 < m_0$, $m_{1/2} \leq 1$ TeV,
$2 \leq \tan\beta \leq 45$, $|A_0| \leq 4 m_{1/2}$.

It has been known from the beginning~\cite{b8,b9} that $a_{\mu}^{\rm SUGRA}$
increases with $\tan \beta$. 
The leading terms come from the chargino ($\Xpm$) diagrams where
the muon can polarize into a chargino and a sneutrino. Expanding for $(\mu \pm 
\tilde{m}_2)^2 \gg M_W^2$ (which generally is the case for mSUGRA), one finds 
for this leading term~\cite{b16},
\begin{equation}
     a_{\mu}^{\rm SUGRA} \cong \frac{\alpha}{4 \pi} \frac{1}{\sin^2 \theta_W}
     \left( \frac{m_{\mu}^2}{m_{\tilde{\chi}^{\pm}_1} \mu} \right) \frac{\tan
     \beta}{1-\frac{\tilde{m}_2^2}{\mu^2}} \left[ 1-\frac{M_W^2}{\mu^2}
    \frac{1+3\frac{\tilde{m}_2^2}{\mu^2}}{\left
(1-\frac{\tilde{m}_2^2}{\mu^2}\right)^2}\right] F(x)
   \label{eq8}
\end{equation}
where $m_{\Xpm} \cong \tilde{m}_2 \cong 0.8 m_{1/2}$,  $x =
m_{\tilde{\nu}}^2/ m_{\Xpm}^2$ and $F(x)$ is a form 
factor arising from the loop integration. This implies that the sign of 
$a_{\mu}^{\rm SUGRA}$ is determined by the sign of $\mu$~\cite{b17}, and since
experimentally, 
Eq. (2), one has that $a_{\mu}$ has a positive anomaly, one has that
\begin{equation}
     \mu > 0		 \label{eq9}
\end{equation}

This result has immediate consequences for dark matter detection. Thus the 
scattering of neutralinos by nuclei in dark matter detectors is governed by 
the cross section $\SXp$, and for $\mu < 0$,
cancellations can occur 
reducing these cross sections  to below $10^{-12}$ pb in large  regions of 
the parameter space~\cite{b18,b19}. This is exhibited  in Fig. 1. If $\mu$ were 
negative, one  would require a detector well beyond anything that is 
currently planned, and so dark matter experiments would not be able to scan 
the full SUSY parameter space. However, with Eq. 9, one finds that cross 
sections generally are greater than $10^{-10}$~\cite{b19}, accessible to future 
planned detectors such as GENIUS~\cite{b20} or Cryoarray~\cite{b21}.

\begin{figure}[htb]
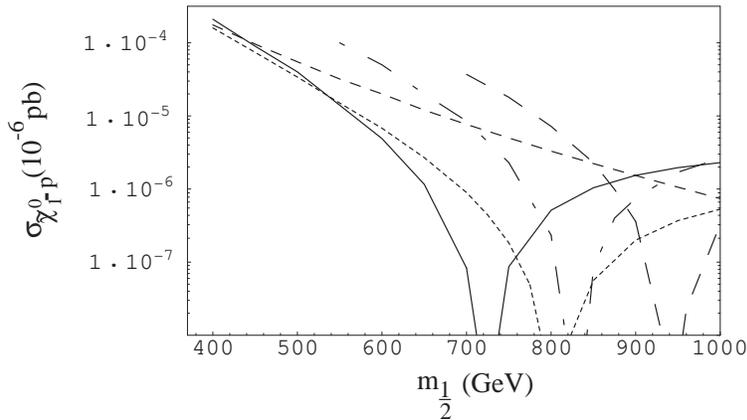

\bigskip
\centerline{ \DESepsf(aadcoan61020.epsf width 10 cm) }
\caption {\it $\sigma_{\tilde{\chi}_{1}^{0}-p}$
for $\mu < 0$, $A_0 = 1500$ GeV, for $\tan\beta = 6$
(short dash), 
$\tan\beta = 8$ (dotted), $\tan\beta = 10$ (solid), $\tan\beta = 20$
(dot-dash), $\tan\beta=25$ (dashed) $^{\displaystyle {19)}}$.}
\label{fig1}
\end{figure}

Eq. (9) also has important theoretical consequences for the mSUGRA model. 
Thus we first note that the combined experimental constraints on the Higgs 
mass and the $\BSG$ branching ratio, implies that most of the 
allowed parameter space is in the co-annihilation region of the parameter 
space, i. e. $m_{1/2} \stackrel{>}{\sim} 300$ GeV. This means that $m_0$ is
essentially determined in 
terms of $m_{1/2}$ (for fixed $A_0$ and $\TB$) as can be seen in Fig. 2. In 
particular, this means that $m_0$ is an increasing function of $m_{1/2}$. Now 
$a_{\mu}^{\rm SUGRA}$ decreases with increasing $m_{1/2}$  and increasing $m_0$.
Thus the lower 
bound of $a_{\mu}^{\rm SUGRA}$ of Eq. (7) then implies an upper bound on
$m_{1/2}$ (since an 
increase in $m_{1/2}$ cannot be compensated by a decrease of $m_0$). Thus the 
combined effects of the data is to give now both lower bounds on 
$m_{1/2}$  (from $m_h$ and $\BSG$) and upper bounds on $m_{1/2}$ (from
$a_{\mu}$), as 
illustrated in Fig. 3. If however, the $a_{\mu}$ data had required  $\mu <0$ 
(instead of Eq. (9)) with the same lower bound on $|a_{\mu}^{\rm SUGRA}|$, then
the $m_h$ and $\BSG$ 
constraints for this sign of $\mu$ would have essentially eliminated all the
mSUGRA parameter space. 
Thus Eq. (7) also represents an experimental test of the 
validity of the mSUGRA model.

\begin{figure}[htb]
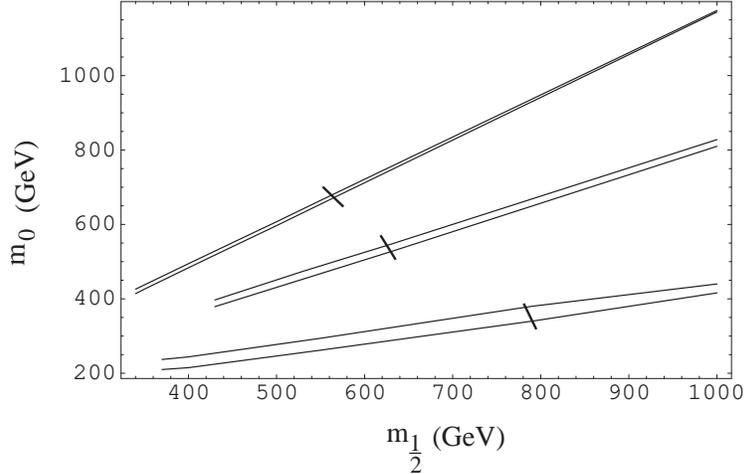

\bigskip
\centerline{ \DESepsf(adhs3.epsf  width 10 cm) }
\caption {\it Corridors in the $m_0 - m_{1/2}$ plane allowed by the
relic density 
constraint for $\tan\beta = 40$, $m_h  > 111$ GeV, $\mu > 0$ for $A_0 = 0, 
-2m_{1/2}, 4m_{1/2}$ from bottom to top. The curves terminate at low $m_{1/2}$
due to 
the $b \rightarrow s\gamma$ constraint except for the $A_0 =4m_{1/2}$ which
terminates due to the $m_h$ constraint. The short  lines through the allowed corridors 
represent the
high $m_{1/2}$ termination due to the lower bound on $a_{\mu}$ of Eq. (7).
$^{\displaystyle {16)}}$}
\label{fig2}
\end{figure}

\begin{figure}[htb]
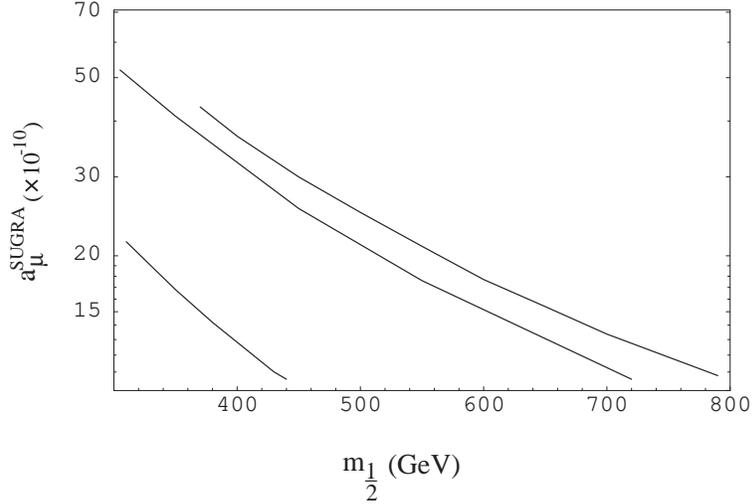

\bigskip
\centerline{ \DESepsf(adhs6.epsf  width 10 cm) }
\caption {\it mSUGRA contribution to $a_{\mu}$ as a function of $m_{1/2}$ for 
$A_0$ = 0, $\mu > 
0$, for  $\tan\beta = 10$, 30 and 40 (bottom to top). $^{\displaystyle{16)}}$}
\label{fig3}
\end{figure}

Since $m_h$ is an increasing function of $m_{1/2}$ and $\tan\beta$, the fact
that $m_{1/2}$ 
is bounded from above, and $m_h$ is bounded from below implies a lower bound 
on $\tan\beta$. We find at the $95\%$ C. L.  that for $m_h > 114$ GeV
\begin{equation}
        \tan\beta > 7 (5) \quad {\rm for} \quad  A_0 = 0 (-4 m_{1/2}) 
	\label{eq10}
\end{equation}
and for $m_h > 120$ GeV one would have
\begin{equation}
        \tan\beta > 15 (10) \quad {\rm for} \quad  A_0 = 0 (-4m_{1/2})
	\label{eq11}
\end{equation}
We see that the $a_{\mu}$ anomaly favors large $\tan\beta$. Fig. 3 also shows
that 
the mSUGRA model cannot accommodate large values of $a_{\mu}^{\rm SUGRA}$, and
if the 
final E821 results were significantly higher than $50 \times 10^{-10}$, this
would 
be a signal for non-universal soft breaking.

The strong constraints on $m_{1/2}$ discussed above also sharpens considerably 
the predictions that mSUGRA has for the SUSY mass spectrum expected at 
accelerators. To see what these are, we consider the $90\%$ C. L. on the
$a_{\mu}$ 
anomaly, i.e. we require now $a_{\mu}^{\rm SUGRA} > 20 \times
10^{-10}$~\cite{b22}.
In this case one 
finds that for $A_0 = 0$ that $\tan\beta > 10$. Then for $\tan\beta < 40$,
$m_{1/2}$ and $m_0$ 
are constrained as follows:
\begin{equation}
      m_{1/2} = (190- 550)\; {\rm GeV};   \quad   m_0 = (70 - 300)\; {\rm GeV}
      \label{eq12}
\end{equation}

In Table 1 we give the range expected for the SUSY masses implied by these 
constraints. All the masses except the light stop, $\tilde{t}_1$, are
insensitive to 
$A_0$ and the lower bound for $m_{\tilde{t}_1}$ can be lowered to 240 GeV by
decreasing 
$A_0$ to $-4m_{1/2}$. From this we see that the trilepton signal unfortunately 
would be out of reach of the Tevatron RUN II ($\tan\beta$ and $m_{1/2}$ are too 
large~\cite{b23}) and RUN II (with $30\; {\rm fb}^{-1}$) would only be able to see the 
light Higgs  provided $m_h < 130$ GeV~\cite{b24}. A 500 GeV NLC would be able
to detect the $h$, 
$\tilde{\tau}_1$, and cover only part of the parameter space for the light
selectron, $\tilde{e}_1$. The LHC, however, would be able to see the full SUSY
spectrum.

\bigskip
Table 1. {\it Allowed ranges for SUSY masses in GeV for mSUGRA assuming 
90\% C. L. 
for  $a_{\mu}$ for $A_0=0$. The lower value of $m_{\tilde t_1}$can be reduced to
240 GeV by
changing $A_0$ to -$4m_{1/2}$. The other masses are not sensitive to
$A_0$.}~\cite{b16}
\begin{center}
 \begin{tabular}{|c|c|c|c|c|c|c|}  \hline
$ \tilde\chi^0_1$ &        $\tilde\chi_1^{\pm}$&$\tilde g$ &       
$\tilde\tau_1$&
       $\tilde  e_1$&         $\tilde  u_1$&         $\tilde  t_1$\\
  \hline
(123-237)&(230-451)&(740-1350)&(134-264)&(145-366)&(660-1220)&(500-940)\\\hline
\end{tabular}
\end{center}
\bigskip

\begin{figure}[htb]
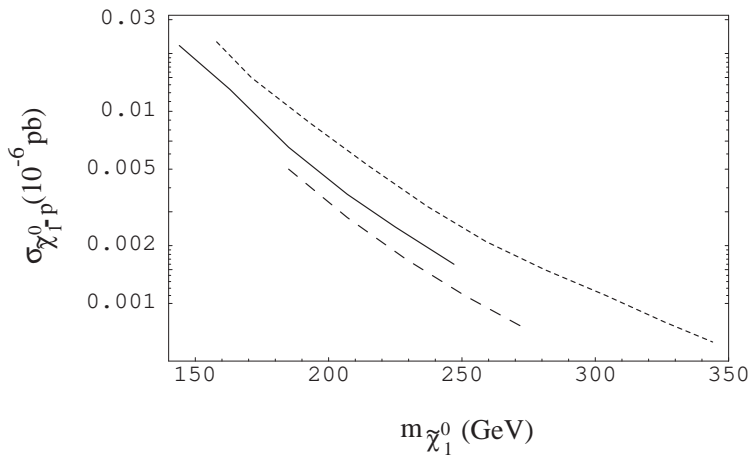

\bigskip
\centerline{ \DESepsf(adhs2.epsf  width 10 cm) }
\caption {\it $\sigma_{\tilde{\chi}_1^0-p}$ as a function of the
neutralino mass
$m_{\tilde{\chi}_1^0}$ for $\tan\beta = 40$, $\mu > 0$ for $A_0 = -2 m_{1/2}, 4
m_{1/2}, 0$ from bottom to top. The curves terminate 
at small $m_{\tilde{\chi}_1^0}$  due to the $b \rightarrow s\gamma$ constraint
for $A_0 = 0$ and $- 2 m_{1/2}$ and 
due to the Higgs mass bound ($m_h > 114$ GeV) for $A_0 = 4 m_{1/2}$. The curves 
terminate at large $m_{\tilde{\chi}_1^0}$ due to the lower bound on $a_{\mu}$
of Eq. (7). $^{\displaystyle{16)}}$}
\label{fig4}
\end{figure}

We saw above that the experimental requirement that $a_{\mu}^{\rm SUGRA}$ be
positive, 
i. e. that $\mu > 0$, eliminated much of the parameter space that would not be 
accessible to dark matter detectors. In addition, the lower bound of Eq. 
(7) produces an upper bound on $m_{1/2}$ and $m_0$, and since the dark matter 
detection cross section, $\SXp$, decreases with increasing $m_0$ and $m_{1/2}$, 
the lower bound on $\SXp$ is raised for the remaining $\mu > 0$ part of the 
parameter space. Fig. 4 shows the expected cross sections for $\tan\beta = 40$, 
$\mu > 0$, $m_h > 114$ GeV, for $A_0 = -2m_{1/2}$, $4m_{1/2}$, 0 (bottom to
top). We see that 
there is a significant dependence on $A_0$, and over the full range, 
$\SXp > 6 \times 10^{-10}$ pb. If the Higgs bound was raised to $m_h > 120$
GeV, then the lower bounds on $m_{1/2}$  increase to 200 GeV, 215 GeV, and 246
GeV 
respectively, significantly further reducing the parameter space. 

If we 
reduce $\tan\beta$ one would expect $\SXp$ to fall. However, for lower 
$\tan\beta$, the upper bound on $m_{1/2}$ becomes more constraining,
eliminating more of the high $m_{1/2}$, $m_0$ region, and thus
compensating partially for the reduction of $\TB$ . This is seen in Fig. 5
where $\SXp$ is 
given for $\tan\beta = 10$, $A_0 = 0$ (upper curve) and $A_0 = -4 m_{1/2}$
(lower 
curve). We have now that $\SXp > 4 \times 10^{-10}$ pb. Thus almost all the 
parameter space should now be accessible to future planned detectors such 
as GENIUS and Cryoarray.

\begin{figure}[htb]
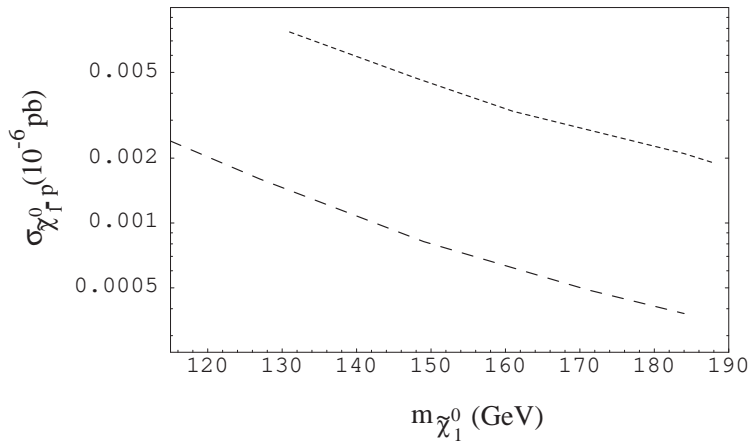

\bigskip
\centerline{ \DESepsf(adhs1.epsf  width 10 cm) }
\caption {\it $\sigma_{\tilde{\chi}_1^0-p}$ as a function of 
$m_{\tilde{\chi}_1^0}$
for $\tan\beta = 10$, $\mu > 0$, $m_h > 114$ 
GeV for $A_0 = 0$ (upper curve), $A_0 = -4 m_{1/2}$ (lower curve). The
termination 
at low $m_{\tilde{\chi}_1^0}$ is due to the $m_h$ bound for $A_0 = 0$, and the
$b \rightarrow  s\gamma$ bound for 
$A_0 = -4 m_{1/2}$. The termination at high $m_{\tilde{\chi}_1^0}$ is due to
the lower bound on $a_{\mu}$ 
of Eq. (7). $^{\displaystyle{16)}}$}
\label{fig5}
\end{figure}


\section{Non-Universal Models}

We consider here the case where there is non-universal soft breaking masses 
at $M_G$ for the Higgs bosons and the third generation of squarks and 
sleptons. This possibility produces interesting new effects. One may 
parameterize the masses in the following way:
\begin{eqnarray} 
m_{H_{1}}^{\ 2}&=&m_{0}^{2}(1+\delta_{1}); 
\quad m_{H_{2}}^{\ 2}=m_{0}^{2}(1+ \delta_{2});\nonumber \\ m_{q_{L}}^{\
2}&=&m_{0}^{2}(1+\delta_{3}); \quad m_{t_{R}}^{\ 2}=m_{0}^{2}(1+\delta_{4});
\quad m_{\tau_{R}}^{\ 2}=m_{0}^{2}(1+\delta_{5});  \nonumber \\ m_{b_{R}}^{\
2}&=&m_{0}^{2}(1+\delta_{6}); \quad m_{l_{L}}^{\ 2}=m_{0}^{2}(1+\delta_{7}).
\label{eq13}
\end{eqnarray} 
where $m_0$ is the universal soft breaking mass for the first two generations, 
$q_L = (\tilde{t}_L, \tilde{b}_L)$, $l_L = (\tilde{\nu}_L, \tilde{\tau}_L)$, and
we assume that $-1 < \delta_i < + 1$. 

\begin{figure}[htb]
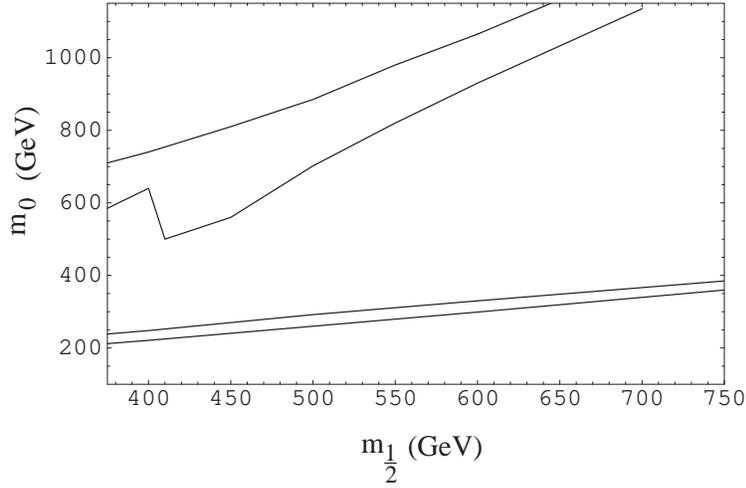

\bigskip
\centerline{ \DESepsf(aadcoan40newnon2.epsf width 10 cm) }
\caption {\it Effect of a non-universal Higgs soft breaking mass
enhancing the $Z^0$
s-channel pole contribution in the early universe annihilation, for the
case of $\delta_2 = $1, $\tan\beta = 40$, $A_0 = m_{1/2}$, $\mu > 0$. The lower
band is
the usual $\tilde\tau_1$ coannihilation region. The upper band is an additional
region satisfying the relic density constraint arising from increased
annihilation via the $Z^0$ pole due to the decrease in $\mu^2$ increasing the
higgsino content of the neutralino. $^{\displaystyle{19)}}$}
\label{fig6}
\end{figure}

While this model has a number of
additional parameters, 
one can understand the physics of what is implied from the following 
considerations. Electroweak symmetry breaking is governed by the $\mu$ 
parameter at the electroweak scale, and for low and intermediate 
$\tan\beta$  one has
\begin{eqnarray}
\mu^2&=&{t^2\over{t^2-1}}\left[({{1-3 D_0}\over 2}+{1\over
t^2})+{{1-D_0}\over2}(\delta_3+\delta_4)\right. \nonumber \\ 
&-&\left.{{1+D_0}\over2}\delta_2+{\delta_1\over
t^2}\right]m_0^2+{\rm {universal\,parts\,+\,loop \, corrections}}. 
\label{eq14}
\end{eqnarray} 
Here $t=\tan\beta$, and $D_0 \cong 1 - (m_t/ 200 \; {\rm GeV} \sin \beta)^2
\cong 0.25$ . ($D_0$ arises 
from running the RGE from $M_G$ to the electroweak scale.) One sees that for 
$\tan\beta > 3$, the universal part of $\mu^2$ depending on $m_0^2$ is quite
small, 
and so this term is sensitive to the amount of non-universal soft breaking 
that might be present. Now much of the physics is governed by $\mu^2$. If
$\mu^2$ 
is decreased, then the higgsino part of $\Xn$ will increase. This has several 
effects. First it increases the $\Xn -\Xn - Z$ coupling which allows then the 
opening of a new neutralino annihilation channel through an s-channel $Z$ pole,
in the relic density 
calculation. This allows a new region of 
allowed relic density at high $\tan\beta$ and high $m_0$. Second, since $\SXp$ 
depends on the interference term between the higgsino and gaugino parts of 
the neutralino, the dark matter detection cross section will be increased 
with increasing higgsino content. To illustrate these effects, we consider 
two cases

\begin{figure}[htb]
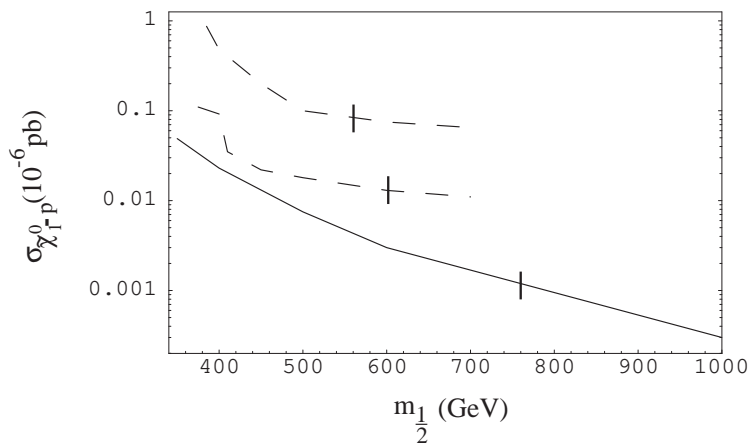

\bigskip
\centerline{ \DESepsf(adhs4.epsf  width 10 cm) }
\caption {\it $\sigma_{\tilde{\chi}_1^0-p}$ as a function of $m_{1/2}$
($m_{\tilde{\chi}_1^0} \stackrel{\sim}{=} 0.4 m_{1/2}$) for $\tan\beta = 40$, 
$\mu >0$, $m_h > 114$ GeV, $A_0 = m_{1/2}$ for $\delta_2 = 1$. The lower curve
is for the $\tilde{\tau}_1-\tilde{\chi}_1^0$ co-annihilation channel, and the
dashed band is for the $Z$ s-channel 
annihilation allowed by non-universal soft breaking. The curves terminate 
at low $m_{1/2}$ due to the $b \rightarrow s\gamma$ constraint. The vertical
lines show the 
termination at high $m_{1/2}$ due to the lower bound on $a_{\mu}$ of Eq.
(7). $^{\displaystyle{16)}}$}
\label{fig7}
\end{figure}

(1) $\delta_2 =1$, $\delta_i = 0$, $i \neq 2$.

Here one sees that the $m_0^2$ term contributes negatively to $\mu^2$.  The 
effect of this is shown in Fig. 6 where the region allowed by the relic 
density constraint in the $m_0 - m_{1/2}$ plane for $\tan\beta = 40$ , $A_0 =
m_{1/2}$ is 
given. The lower corridor, coming from the stau-neutralino co-annihilation 
is much the same as the $A_0 = 0$ corridor in Fig. 2. The upper channel, at 
higher $m_0$, is due to the s-channel $Z$- pole annihilation. Fig. 7 shows the 
neutralino proton cross sections for corresponding allowed regions as a 
function of $m_{1/2}$ ($m_{\Xn} \cong 0.4 m_{1/2}$), for $\tan\beta = 40$. In
spite of the fact 
that $m_0$ is quite high for the $Z$ s-channel corridor, the cross sections is 
quite large, and at least part of the this region should be accessible to 
GENIUS -TF, and CDMS in the Soudan mine.  This increase is due to the fact 
that the non-universality has lowered $\mu^2$.

\begin{figure}[htb]
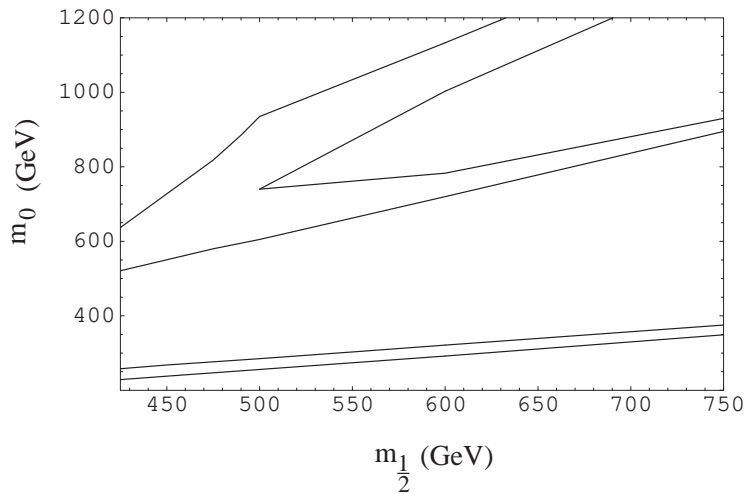

\bigskip
\centerline{ \DESepsf(aadcoan40newnon.epsf width 10 cm) }
\caption {\it Allowed regions in the $m_0-m_{1/2}$ plane for the case 
$\tan\beta = 40$, $A_0
= m_{1/2}$, $\mu > 0$. The bottom curve is the mSUGRA $\tilde\tau_1$ 
coannihilation band of
Fig. 2 (shown for reference). The middle band is the actual $\tilde\tau_1$
coannihilation band when $\delta_{10} = -0.7$. The top band is an additional
allowed region due to the enhancement of the $Z^0$ s-channel annihilation
arising from the nonuniversality lowering the value of $\mu^2$ and hence 
raising the higgsino content of the neutralino. 
For $m_{1/2}\stackrel{<}{\sim}$ 500 GeV, the
two bands overlap. $^{\displaystyle{19)}}$}
\label{fig8}
\end{figure}

(2) $\delta_{10} (= \delta_3  = \delta_4 = \delta_5) = -0.7$.

This is a model that might arise in a $SU(5)$ GUT, where only the particles 
of the $\mathbf{\overline{10}}$ representation 
in the third generation have non-universal soft 
breaking masses. Again, the non-universalities lower the value of $\mu^2$, 
allowing for a $Z$ s-channel corridor in the relic density analysis. In 
addition, because the soft breaking mass of the stau is reduced, the 
co-annihilation channel occurs at a higher value of $m_0$. This is shown in 
Fig. 8, for $\tan\beta = 40$, where the bottom band, the usual mSUGRA 
co-annihilation channel, is now moved up to much higher $m_0$. (This 
co-annihilation and the $Z$-channel annihilation corridors merge for $m_{1/2}
\stackrel{<}{\sim} 500$ GeV.)  The corresponding cross sections are shown in
Fig. 9. Again part 
of this parameter space should be accessible to the next round of dark 
matter detectors.

\begin{figure}[htb]
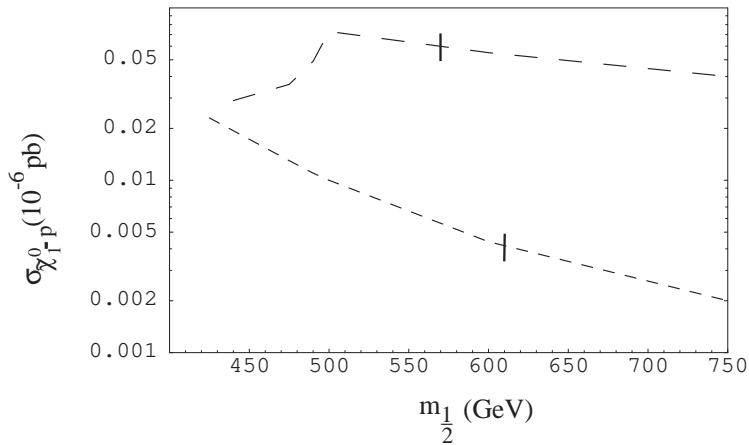

\centerline{ \DESepsf(adhs5.epsf  width 10 cm) }
\caption {\it $\sigma_{\tilde{\chi}_1^0-p}$ as a function of $m_{1/2}$ for 
$\tan\beta = 40$, $\mu >0$, $A_0 = m_{1/2}$ 
and $m_h  > 114$ GeV. The lower curve is for the bottom of the
$\tilde{\tau}_1-\tilde{\chi}_1^0$ 
co-annihilation corridor, and the upper curve is for the top of the $Z$ 
channel band. The termination at low $m_{1/2}$ is due to the $b \rightarrow
s\gamma$ 
constraint, and the vertical lines are the upper bound on  $m_{1/2}$ due to the 
lower bound of $a_{\mu}$ of Eq. (7). $^{\displaystyle{16)}}$}
\label{fig9}
\end{figure}

\section{Conclusions}

We have examined here the $2.6\,\sigma$ deviation of $a_{\mu} = (g_{\mu}
-2)/2$ from its 
predicted Standard Model result under the assumption that the effect is 
real and can be understood in terms supergravity GUT models with R-parity 
invariance. For mSUGRA models, the combined constraints from the $a_{\mu}$,
$m_h$, 
and the  $\BSG$ decay constraints and the relic density of $\Xn$ dark 
matter greatly limit the SUSY parameter space, greatly tightening the 
predictions the theory can make. The lower bound on $a_{\mu}$ produces an
upper 
bound on $m_{1/2}$, and then the $m_h$ and $\BSG$ constraints produce lower 
bounds on $m_{1/2}$ and $\tan\beta$. We find that for $m_h > 114$ GeV, that
$\tan\beta  > 
7(5)$ for $A_0 = 0 (-4m_{1/2})$ and for $m_h > 120$ GeV, $\tan\beta > 15 (10)$
for $A_0 = 0 
(-4m_{1/2})$. The lower bound on $m_{1/2}$ pushes most of the parameter space
into 
the co-annihilation domain effectively determining $m_0$ in terms of $m_{1/2}$. 
However, there is still a significant dependence on $A_0$ and $\tan\beta$.

The restrictions on the allowed range of $m_{1/2}$ and $m_0$ allow one now to
make 
significant predictions concerning accelerator reaches for the SUSY 
particles. Thus we find at $90 \%$ C. L. bounds on $a_{\mu}$ that the Tevatron
Run 
II should be able to see the light Higgs (if $m_h < 130$ GeV), but no other 
parts of the SUSY spectrum. A 500 GeV NLC should be able to see the $h$ and 
$\tilde{\tau}_1$, and scan part of the parameter space where the $\tilde{e}_1$
is expected. The 
LHC should be able to detect the full SUSY spectrum. Further, future 
planned dark matter detectors, GENIUS and Cryoarray, should be able to 
sample almost all of the SUSY parameter space. For other analyses within the
mSUGRA framework see~\cite{b25}.

Non-universal SUGRA models can allow new regions in parameter space to open 
due to possible annihilation in the early universe through s-channel $Z$ 
poles. For these types of non-universalities, which lower the value of 
$\mu^2$, the neutralino proton cross sections become larger, and part of these 
effects can possibly be tested with current detectors. Since 
non-universalities are phenomena determined by post-GUT physics, the dark 
matter detectors could allow one to learn about physics beyond the GUT scale.

The current anomaly in $a_{\mu}$ is a $2.6\,\sigma$ effect. However, the
additional 
data that now exists, and the expected reduction of the theoretical errors 
should allow a determination in the near future of whether the effect is 
real.

\section{Acknowledgements}
This work was supported in part by National Science Foundation grant No.
PHY-0070964.
%

%
\end{document}